\documentclass[a4paper]{article}
\usepackage[margin=25mm]{geometry}
\usepackage{amsmath}
\usepackage{amsfonts}
\usepackage{amssymb}
\usepackage{graphicx}
\usepackage{verbatim}
\immediate\write18{texcount -tex -sum  \jobname.tex > \jobname.wordcount.tex}

\providecommand{\keywords}[1]
{
  \small	
  \textbf{\textit{Keywords---}} #1
}

\title{\huge  A Parametric Approach to Relaxing the Independence Assumption in Relative Survival Analysis}

\author{Adatorwovor, R$^{1}$, Latouche,A.$^{2}$, Fine, J., P.$^{3}$ \\
        \small $^{1}$University of Kentucky, radatorwovor@uky.edu \\
        \small $^{2}$University Conservatoire National des Arts et Métiers, Paris, France; \\    \small Institut Curie, St-Cloud, France, aurelien.latouche@curie.fr \\
        \small $^{3}$University of North Carolina at Chapel Hill, jpfine@email.unc.edu\\
}
\date{} 

\begin{document}
\maketitle

\begin{abstract}
With known cause of death (CoD), competing risk survival methods are applicable in estimating disease-specific survival. Relative survival analysis may be used to estimate disease-specific survival when cause of death is either unknown or subject to misspecification and not reliable for practical usage. This method is popular for population-based cancer survival studies using registry data and does not require CoD information. The standard estimator is the ratio of all-cause survival in the cancer cohort group to the known expected survival from a  general 
reference population. Disease-specific  death competes with other causes of mortality, potentially creating dependence among the CoD. The standard ratio estimate is only valid when death from disease and death from other causes are independent. To relax the independence assumption, we formulate dependence using a copula-based model. Likelihood-based parametric method is used to fit the distribution of disease-specific death without CoD information, where the copula is assumed known and the distribution of other cause of mortality is derived from the reference population. We propose a sensitivity analysis, where the analysis is conducted across a range of assumed dependence structures. We demonstrate the utility of our method through simulation studies and an application to French breast cancer data.
\end{abstract} \hspace{10pt}


\verbatiminput{\jobname.wordcount.tex}

\section{Introduction}
\label{s:intro}
Cancer patients including breast, prostate, endometrial and thyroid cancer are at higher risk of dying from heart disease and stroke than the general population. As the number of cancer survivors increases, so is the rate of cardiovascular deaths (Sturgeon {\it{ et al.}},  2019).  A patient's survival burden can be quantified either by the distribution of cancer-specific death in the presence of death from other causes  or by the cancer-specific mortality in the absence of failure types other than the disease of interest. This quantity, also known as net survival (under independence assumption), is controversial but meaningful to many practitioners and researchers for comparisons of survival across populations with different background mortality.

With improvement in medical treatment and long follow-up in population-based disease registries, patients may either experience disease-specific death  or death from non-disease related causes (Brinkhof {\it et al.}, 2010). 
In such competing risk settings where one death type precludes the occurrence of other types, standard methodology assumes that cause of death is known (Gichangi and Vach, 2005). 
In the analysis of competing risks events from registry data, accurate documentation of death is essential (Percy{ \it  et al.}, 1981; Welch and  Black, 2002; and Mieno { \it et al.}, 2016). 
A challenge is that documentation either may not be available, or may be incomplete or incorrect for cause of death, resulting in problems distinguishing disease and non-disease related mortality. The issue is pronounced in Europe, where comparison of disease-specific survival across countries is of interest. The World Health Organization (World Health Organization, 1977)  
defines cause of death as "the disease or injury which initiated the train of morbid events leading directly to death". However, population-based disease registries may not be harmonized across countries, leading to imprecise cause of death definitions and different levels of documentation for cause of death information. Often, the underlying cause of death may be unclear as hospital coding of cancer death may not agree with the death certificate coding. As an example, Welch and  Black (2002) 
reported that $41\%$ of deaths that occurred (within one month of cancer diagnosis and cancer directed surgery) were not attributable to the coded cancer in the registry. When reliable cause of death information is available, it is often located in separate databases, which may be costly to obtain and difficult to link with registry data.

Suppose that $T = \min\{T_k:\ k= 1,2,3,\cdots,K\}$ is the potentially observable failure time  and $\varepsilon=\{k: T = T_k \}$ the failure type where $T_1, T_2, \cdots,T_K$, are the latent failure types associated with K failure types. In registry data, $K = 2$ and $\varepsilon = 1$ implies death from cancer and $\varepsilon = 2$ implies death from other competing causes. Standard methods for independently right censored survival data without competing risks cannot generally be used to make inference about disease-specific survival. Under dependent competing risks, where $T_1$ and $T_2$ are dependent, the Kaplan-Meier (Kaplan and Meier, 1958)  
curve estimates a function of the cause-specific hazard function, defined in Section $2$. The logrank test  (Bland and Altman, 2004) 
assesses group differences between the cause-specific hazard function, while the standard proportional hazards model (Cox, 1972) 
formulates the effects of covariates on the cause-specific hazard function. The cumulative incidence function, defined in Section $2$, gives disease-specific survival in the presence of competing events. This quantity has been widely adopted in applications, with the Aalen-Johanson estimator  (Aalen and  Johansen, 1978), 
 Gray's test (Gray, 1988), 
and the Fine-Gray model (Fine and Gray, 1999), 
providing analogs to the Kaplan-Meier curve, the logrank test, and the proportional hazards model for the cumulative incidence function. Without cause of death information, these methods are not applicable.

To address disease-specific survival without reliable cause of death information, relative survival methods have been proposed. Relative survival, $S_R(t)$, is the ratio of the observed survival rate in a group of cancer patients, during a specified period, to the expected survival rate in a 
general reference population, (Ederer, 1961). 
Mathematically,  
\begin{eqnarray} \label{eq:1}
S_R(t) = \frac{S_O(t)}{S_P(t)}  
\end{eqnarray}
where at time $t$, $S_O(t)$ is the observed survival probability from the registry and $S_P(t) $ is the expected survival from mortality tables. Existing literature has focused exclusively on the estimation of $S_R(t)$ under the independence assumption, $T_1 \perp T_2$. Under independence, $S_O(t) = S_{T_1}(t) \cdot S_{T_2}(t)$, with $\ S_P(t) = S_{T_2}(t)$ which implies $S_R(t) = S_{T_1}(t)$ where $S_{T_1}(t)$ and $S_{T_2}(t) $ are the survival probabilities corresponding to $T_1$ and $ T_2$ respectively. The relationship $(\ref{eq:1})$ can be rewritten in terms of hazard functions as $\lambda_O(t) = \lambda_E(t) + \lambda_P(t)$ (Cronin and Feuer, 2000), 
where $\lambda_O(t)$ is the hazard in the disease registry, $\lambda_E(t)$ is the so called excess hazard among the cancer cohort, and $\lambda_P(t) $ is the hazard from mortality tables. Under independence, $\lambda_E(t)=\lambda_{T_1}(t)$ and $\lambda_P(t)=\lambda_{T_2}(t)$, where $\lambda_{T_j}(t)=\frac{-dlog{S_{T_j}(t)}}{dt}, j=1,2,$ are the net hazard functions for cancer and other cause mortality, respectively. The net survival probability (under independence assumption)  $S_{T_1}(t)$  is the target of relative survival analysis and corresponds to a hypothetical population in which death from competing causes does not exist. It differs from the cumulative incidence function which is commonly used to quantify disease-specific survival in analyses with known cause of death information. 
Under dependence, $S_R(t)$ in  (\ref{eq:1}) has an excess hazard 
(Suissa, 1999) interpretation since the estimator eliminates the effect of background risk, and is no longer the survival probability $S_{T_1}(t)$. 

Relative survival method was pioneered by Berkson and Gage (1950), 
and Ederer { \it et al.} (1961)
for nonparametric estimation of $S_{T_1}(t)$. A variant of this method was proposed by Hakulinen (1982) 
to address the bias due to heterogeneity of patient withdrawal within subgroups. Pohar Perme et al., (2012) 
demonstrated that these classical methods may be biased under certain censoring patterns. For example, in population comparisons, such bias may arise from unmeasured covariates affecting the cancer cohort group and the reference population from which rates of expected mortality are drawn. Rebolj Kodre and  Pohar Perme, (2013) 
studied biases associated with censoring and age distribution (at the time of cancer diagnosis) and proposed weighting corrections. 
Nixon et al. (1994) documented that event times and censoring times are dependent on the age of the patients in a cancer study. Stratified methods (Sasieni and Brentnall, 2017)
based on age standardization of relative survival ratios may reduce such biases. Hakulinen et al., (2011) 
developed  alternative estimators valid under weaker assumption. 

Several authors (Bolard, et al., 2002; Giorgi, et al. 2003; Nelson et al., 2007; Mahboubi, et al., 2011) proposed variations of spline functions to address some of limitations of  earlier methods. These models include mainly flexible parametric functions that examine the effects of covariates that potentially influence the estimation of the excess survival.  Charvat et al., (2016) introduced spline models for grouped data, where they fitted flexible excess hazard models to cluster data, while Rubio et al., (2019) proposed extensions (beyond proportional hazard method) to the general parametrization of hazard functions and implemented flexible parametric distribution in modelling the event time. However, the above estimation methods for $S_R(t)$ do not model the dependence event times concurrently.

To relax the independence assumption,  
we formulate the dependence between the latent failure times distributions for death from disease and death from competing causes using copula models (Deheuvels, 1978). 
A copula function generates a joint distribution for the two event times, taking as input their marginal distributions. Copulas allow a broad range of dependence structures and have been employed widely in survival analysis, including bivariate event times (Oakes, 1982), 
competing risks with known cause of failure (Heckman and Honoré, 1989), 
and semi-competing risks where one event time censors the other but not vice versa (Fine {\it et al.}, 2001). 
We employ such models with competing risks data from disease registries where cause of death information is either not reliable or not available. Because the joint distribution of the latent failure times is nonparametrically nonidentifiable (Tsiatis, 1975), 
we treat the copula function as known. The marginal distribution of the time to disease-specific death is modelled parametrically with the distribution of death from other causes drawn from the reference population. Likelihood-based inference is proposed. Because the joint distribution is unidentifiable nonparametrically and unverifiable from the observed registry data, a sensitivity analysis is suggested in which disease-specific survival is estimated across a range of rich dependence structures, specified via the copula function. To our knowledge, this is the first attempt in modelling dependence between $T_1 $ and $T_2$ in relative survival analysis.

The main purpose of this method is to provide an alternative estimator for net survival (survival in a hypothetical world where other competing causes of death do not exist) under dependence between cancer mortality and other cause mortality. This is accomplished by explicitly modelling the dependence between cancer mortality and other cause mortality.  The rest of this paper proceeds as follows. In section $2$, we present the data and copula model formulation for competing risks data. Section $3$ describes the likelihood estimation and inference procedure without cause of death information, as well as the proposed sensitivity analysis. In section $4$, we present the numerical illustrations including simulation results and application to French breast cancer data. Section $5$ discusses and concludes the paper.

\section{Data and Model}
\label{s:data}
We begin by defining traditional endpoints for competing risk data with known cause of death. The cause-specific hazard, $\lambda_k(t)$ is the instantaneous failure rate for occurrence of event $\varepsilon = k$  at time t (Prentice et al., 1978), 
 \begin{eqnarray} \label{eq:2}
 \lambda_k(t)= \lim_{\delta t \rightarrow 0} \frac{P(t \leq T < t + \delta t, K = k| T> t)}{\delta t}
 \end{eqnarray}
 and the cumulative incidence function $C_k(t)$ is the proportion of patients who died from cause $k$ by time t in the presence of patients who might die from other causes. The  cumulative incidence function 
 can be expressed as $C_k(t) =  P (T \leq  t, \  \varepsilon = k) = \int^t_0 \lambda_k(s) \cdot S(s)ds$ where $S(t) = P(T > t)$ is the overall survival probability. Standard competing risks methods with known cause of failure focus on estimation of $\lambda_k(t) $ and $ C_k(t)$. 
 
Without cause of death information, the registry data is simply time to death from any cause, T, which may be right censored by lost to follow up. Let C be the time to right censoring, with the common assumption being that T and C are independent. The observed data consist of $X_i=\min(T_i,C_i)$ and $\delta_i=I(T_i\leq C_i)$, where $T_i$ and $C_i$ are the failure and censoring times on individual $i=1,2,3,\cdots,n$. Relative survival methods employing such data do not focus on the traditional competing risks endpoints $\lambda_k(t)$ and $C_k(t)$ but rather on the latent failure time distributions with the corresponding survival functions $S_{T_1}(t)$ and $S_{T_2}(t)$. 

To capture the dependence between $T_1$ and $T_2$, we employ copula models, which completely describe the dependence structure and provide scale invariant measures of association (Venter, 2002; Müller, 1996; Bäuerle and Müller, 1998; and Denuit {\it et al.}, 1999).  
Suppose $\psi$ is a copula generator function defined such that $\psi:[0,1] \rightarrow [0,+ \infty]$,  
  then the j-dimensional copula function is given by $C(u_1,\cdots,u_j) =   \psi \left(\psi^{-1} (u_1),\cdots, \psi^{-1} (u_j) \right) $ with marginal distributions, $u_j = P(T_j\leq t_j) = F_{T_j}(t_j) =1 - S_{T_j}(t_j),  \ \forall j \in  \mathbb{N}$. When $j= 2$, then, the copula model for the joint distributions of $T_1$ and $T_2$ is:
\begin{eqnarray*}
 C(u_1,u_2)  = P(T_1\leq t_1,T_2 \leq t_2) =
 \psi \left(\psi^{-1} (u_1) +\psi^{-1} (u_2) \right)
 = F_{T_1,T_2}(t_1,t_2)
 \end{eqnarray*}
 where  
 $\psi^{-1}$ is the inverse of $\psi$ and $\psi$ satisfies the Laplace-Stiltjes transform and Bernstein (1929) theorem.  
 McNeil and Ne{\v{s}}lehov{\'a} (2009) 
 showed that the generator function $\psi$ is completely monotone for non-negative random variables with $ \psi(0) = 1, \  \psi'(\cdot) < 0 $ and $\psi''(\cdot) <0$.

The most widely used scale invariant measures of association to characterize dependence are Spearman's rho ($\rho_S$) and Kendall's tau ($\tau_{ken}$) correlation coefficients. The connection between the latter and the copula generator function was shown by Genest and  MacKay, (1986) as: 
 \begin{eqnarray*}
 \tau_{ken} = 1+4\int_0^1 \frac{\psi^{-1}(u)}{\psi^{-1}(u)'}du = 1 -4\int_0^\infty u(\psi(u))^2 du 
 \end{eqnarray*}
 with  $\psi^{-1'}$ being the derivative of $\psi^{-1}$.
 While in theory, any copula may be used to link the marginal distributions of $T_1$ and $T_2$, in this paper, we focus on two popular Archimedean copulas, indexed by a single dependence parameter $\theta$ having simple interpretations. The Gumbel copula is:
 \begin{eqnarray} 
 C(u_1,u_2)= \exp\left[- \{(-log(u_1))^\theta + (-log(u_2))^\theta\}^{\frac{1}{\theta}} \right]
 \end{eqnarray}
 with $\theta \in (1,+\infty)$  
 and the Clayton copula is:
 \begin{eqnarray}\label{eq:4}
 C(u_1,u_2) = (u_1^{-\theta} + u_2^{-\theta}-1)^{-\frac{1}{\theta}}
 \end{eqnarray} 
 with $\theta \in (0,+\infty)$. 
 A special case of product copula model: $C(u_1,u_2)=u_1 \cdot u_2$ is obtained when $\theta = 1$ and when $\theta \rightarrow 0$ for Gumbel and Clayton copulas respectively. The product copula model gives independence of $T_1$ and $T_2$. When $\theta > 0$, the Clayton copula is bounded by: $C(u_1,u_2)\leq \theta(1-u_1-u_2)+(1+\theta)u_1u_2$. As dependence increases, that is  $\theta \rightarrow +\infty$, the Clayton copula approximates the Fre{\'chet-Hoeffding} upper bound, (Fréchet, 1951; and Hoeffding, 1940) giving perfect positive dependence. 

\section{Likelihood Estimation and Inference}
\label{s:like}
We first formulate our model without covariates for the potentially dependent latent failure times $T_1$ and $T_2$. In the sequel, the distribution of $T_1$ involves an unknown parameter $\eta$, while that of $T_2$ is assumed from a reference population. The distribution of T implicitly involves $\eta$ via $T = \min(T_1,T_2)$. The survival function for all-cause mortality at  time 
t given $\eta$ is:
\begin{eqnarray} \label{eq:5}
S_{T}(t|\eta)  & = & S_{T_{1}} (t|\eta) + S_{T_{2}} (t) - 1 + F_{T_{1},T_{2}}(t,t|\eta) \nonumber \\ 
& = & 1 -  F_{T_{1}}(t|\eta) - F_{T_{2}}(t) + F_{T_{1},T_{2}}(t,t|\eta)
\end{eqnarray}
with the corresponding density function using $F_{T}(t|\eta)  =  F_{T_{1}}(t|\eta) + F_{T_{2}}(t) -  F_{T_{1},T_{2}}(t,t|\eta) $) 
\begin{eqnarray} \label{eq:6}
f_{T}(t|\eta)   =  
f_{T_{1}}(t|\eta) + f_{T_{2}}(t) - f_{T_{1},T_{2}}\left(t,t|\eta \right)
\end{eqnarray}  
where $f_{T_1} (t|\eta) = \frac{dF_{T_j}(t|\eta)}{dt}$, $f_{T_1,T_2}(t,t|\eta) = \frac{d F_{T_1,T_2}(t,t|\eta)}{dt}$ and $f_{T_2}(t)$ is derived from the reference population. 
If censoring of $T$ by $C$ is noninformative, then the likelihood contribution for individual i is:
\begin{eqnarray} \label{eq:7}
L_i = f_{X_i,\Delta_i}(X_i,\delta_i|\eta) = [f_{T}(X_i|\eta)]^{\delta_i} [S_{T}(X_i|\eta)]^{1 - \delta_i} 
\end{eqnarray}
From equation $(\ref{eq:7})$, the full log-likelihood function based on n independent observations is:
\begin{eqnarray} \label{eq:8}
l(\mathbf{X},\Delta|\eta) & = &\sum_{i = 1}^n \left( \delta_i*\log f_{T}(X_i|\eta) + (1 - \delta_i)*\log S_{T}(X_i|\eta) \right) \nonumber \\
& = & \sum_{i=1}^n  \delta_i * \log\left[f_{T_{1}}(X_i|\eta) + f_{T_{2}}(X_i) - f_{T_{1},T_{2}}\left(X_i,X_i|\eta \right)\right]\nonumber \\
& & + \sum_{i = 1}^n (1 -  \delta_i ) *\log\left[S_{T_{1}} (X_i|\eta) + S_{T_{2}} (X_i) -1 + F_{T_{1},T_{2}}(X_i,X_i|\eta) \right] 
\end{eqnarray}
where $(\mathbf{X},\Delta) = (X_i,\Delta_i, i = 1,2,3,\cdots,n)$. 

We specify a parametric model for $F_{T_1}(t)$, with finite dimensional parameter of interest $\eta$. The general form of the probability density function for $T_1$ at time t is $f_{T_1}(t|\eta)$ with survival probability $S_{T_1}(t|\eta) = 1 -  F_{T_1} (t|\eta)= \int_t^\infty f_{T_1}(s|\eta)ds $. The distribution of $T_2$ is assumed known and extracted from the reference population with the usual assumption that disease-specific death is negligible in this reference population 
(Ederer, et al. 1961). 
This is illustrated in the French breast cancer data analysis in section $4.2$. The copula distribution linking $F_{T_1}(t|\eta)$ and $F_{T_2}(t)$ may be specified using simple parametric copula models such as the Archemedean copulas. The parameters in the copula model may be chosen for a pre-specified dependence between $T_1$ and $T_2$, for example, Kendall's tau ($\tau_{ken}$). In the numerical illustrations, $T_1$ was assumed to follow an exponentiated Weibull distribution with parameter $\eta = (\lambda, \kappa, \alpha)$ and probability density function $f(t: \lambda, \kappa, \alpha) = \alpha \frac{\kappa}{\lambda} \cdot  \left(\frac{t}{\lambda} \right)^{\kappa - 1} \cdot \exp \left\{-  \left(\frac{t}{\lambda} \right)^\kappa  \right\}  \left( 1 -  \exp \left\{-  \left(\frac{t}{\lambda} \right)^\kappa  \right\}     \right)^{\alpha-1} $
because of its versatility to accommodate a wide range of hazard shapes. We consider the Gumbel and Clayton copulas for the joint distribution of $T_1$ and $T_2$ as both copulas exhibit tail behaviours that mimic the mortality trend observed in the cancer registry data. 
The bivariate joint distribution and density functions under the Gumbel copula are given below.

\begin{eqnarray}\label{eq:9}
F_{T_1 ,T_2}(t|\eta,t) & = & \exp\left\{-\left((-log\left(F_{T_1}(t|\eta)) \right)^\theta + \left(-log(F_{T_2}(t) ) \right)^\theta  \right)^{\frac{1}{\theta}}\right\} \nonumber \\
f_{T_{1},T_{2}}(t|\eta,t) & = & F_{T_1,T_2}(t|\eta,t) \cdot \left( \left(-\log\left(F_{T_1}(t|\eta)\right)^\theta \right) + \left(-log \left(F_{T_2}(t) \right)^\theta \right) \right)^{\frac{1}{\theta} -1}
\nonumber \\
&\times & \left( \left(-\log (F_{T_1}(t |\eta) )^{\theta - 1}\cdot\frac{f_{T_{1}}(t|\eta)}{F_{T_1}(t|\eta)} \right)  +  \left(-log \left(F_{T_2}(t|\eta)\right)^{\theta - 1}\cdot\frac{f_{T_{2}}(t)}{F_{T_2}(t)} \right) \right) \nonumber\\ 
 \end{eqnarray}
Under the Clayton copula, the bivariate joint distribution and density functions are: 
\begin{eqnarray} \label{eq:10}
F_{T_{1},T_{2}}(t|\eta,t)
& = &\left( F_{T_1}(t|\eta)^{-\theta } + F_{T_2}(t)^{-\theta } -1\right)^{-\frac{1}{\theta } } \nonumber\\ 
f_{T_{1},T_{2}}(t|\eta,t)& = &\frac{F_{T_{1},T_{2}}(t|\eta,t) } {\left(F_{T_1}(t|\eta)^{-\theta} + F_{T_2}(t)^{-\theta} - 1\right)} \cdot \left( \frac{f_{T_{1}}(t|\eta)}{F_{T_1}(t|\eta)^{\theta + 1}}+ \frac{f_{T_{2}}(t)}{F_{T_2}(t)^{\theta + 1}}  \right)
\end{eqnarray}

The maximum likelihood estimator (MLE) of $\eta$ can be obtained by maximizing the log-likelihood function in $(\ref{eq:8})$ using any optimization algorithm such as  Nelder-Mead  (Nelder and Mead, 1965) or better. 
In the simulation in section 4.1, and because the model is highly nonlinear  for hazard shapes and  the estimation of dependent parameter $\theta$, 
we suggest using multiple starting values and taking the MLE to be the maximizer giving the largest value of the log likelihood across all starting values. Under a fixed and correctly specified copula, the usual regularity conditions for the MLE holds and the estimator converges in probability, that is $\hat{\eta} \xrightarrow{P} \eta $ and is asymptotically normal, 
$\displaystyle{ \hat{\eta} \sim N \left(\eta, I_O(\eta)^{-1}\right)}$  with variance estimated using the inverse of the observed information matrix \begin{eqnarray}
I_O(\hat{\eta}) & = &
-\frac{\partial^2 l(\mathbf{X},\Delta|\mathbf{\eta})}{\partial \eta \partial \eta^T} |_{\eta = \hat{\eta}} \nonumber \\ 
& = &   -\Biggl\{ \delta_i \cdot \sum_{i = 1}^n \left\{ \frac{ 
[ f_{T}(X_i|\eta)]
\cdot \left[\frac{\partial^2}{\partial \eta^2}f_T(X_i|\eta) \right]   - \left[ \frac{ \partial }{ \partial \eta  } f_T(X_i|\eta) \right]^T\left[ \frac{ \partial }{ \partial \eta  } f_T(X_i|\eta) \right] }
{[ f_T(X_i|\eta) ]^2
}  \right\} \nonumber\\ 
& & + (1 - \delta_i ) \cdot \sum_{i = 1}^n \left\{ \frac{ 
[ S_{T}(X_i|\eta)]
\cdot \left[\frac{\partial^2}{\partial \eta^2} S_T(X_i|\eta) \right]   - \left[ \frac{ \partial }{ \partial \eta  } S_T(X_i|\eta) \right]^T\left[ \frac{ \partial }{ \partial \eta  } S_T(X_i|\eta) \right] }
{[ S_T(X_i|\eta) ]^2
}  \right\}  \Biggl\} |_{\eta = \hat{\eta}} \nonumber \\
\end{eqnarray}

Since the dependence structure for time  to disease-specific mortality ($T_1$) and time  to other competing mortality  ($T_2$) is nonidentifiable and unverifiable from the observed registry data, we propose a sensitivity analysis, where the analysis is conducted across a range of assumed dependence structures. The levels of dependence represent the varying levels of dependent competing mortality possible in the observed registry data. For each copula dependence structure with known  $\theta $, we estimate $\eta$  with $\hat{\eta}$ and compute $F_{T_1}(t|\hat{\eta})$ to estimate relative survival. The corresponding standard errors are obtained as the square root of the Delta method variance: $Var(\widehat{S_{T_1}(X)}) = g(\widehat{S_{T_1}(X)}) \cdot I_O(\hat{\eta})^{-1}\cdot g^T(\widehat{S_{T_1}(X)})$ where $g(\eta)$ is the derivative of $S_{T_1}(t|\eta)$ with respect to $\eta$. Due to the complex nature of the likelihood, numerical approximation is used to estimate the information matrix in the numerical illustrations in Section 4. 

In the presence of informative censorship where T and C are dependent, we propose conditioning on additional covariates Z in $F_{T_2}$, 
(Sasieni and Brentnall, 2017; and  Pohar Perme et al., 2012), 
where $F_{T_2}(t|Z)$ is the conditional distribution of $T_2$ given Z. Such covariates might include age, sex, period, as well as other relevant demographic variables. Let $Z_i$ be the covariate observed on individual $i=1,\cdots, n$. The log-likelihood function $(\ref{eq:8})$ is easily modified, where the likelihood contribution for individual i $(=1,\cdots ,n)$ is $(\ref{eq:7})$ with $F_{T_2}(t|Z_i)$ replacing $F_{T_2}(t)$ in $f_T(X_i|\eta)$ and $S_T(X_i|\eta)$. Here, we estimate $\eta$ in $F_{T_1}(t|\eta)$ conditionally on Z to mitigate against the bias (caused by informative censoring) associated with these covariates  
(Pohar Perme et al, 2012; and Sasieni and Brentnall, 2017, Schaffar, et al., 2017). 
The usual likelihood regularity conditions continue to hold, with the resulting estimator $\hat{\eta}$ being consistent and asymptotically normal with variance which may be estimated using the inverse of the observed information matrix evaluated at $\hat{\eta}$.

\section{Numerical Illustrations}
\label{s:num}
\subsection{Simulation Studies}
To evaluate the performance of our proposed method, first, we simulated a  general survival time data $T_j \sim ExpWeibull (\lambda_j,\kappa_j, \alpha_j)$, and second, to mimic the French breast cancer data set for sample sizes; $1000, \ 2500$ and $5000$ with $500$ replications.  The latent failure times for  $T_j$
with probability density function defined above in Section $3$ and parameters as in figure $1$. The exponentiated Weibull distribution degenerates to a 2-parameter Weibull distribution when $\alpha= 1$. The parameters for the reparametrized Weibull distribution for $T_1$ were $ \lambda_1 = 0.182$ and  $\alpha_1 = 1.609$, while  those for  $T_2$ were $\lambda_2 = 0.742 $ and $\alpha_2 = 0.693 $. In the estimation of $\lambda_1, \ \alpha_1$ for $T_1$, $\lambda_2, \ \alpha_2$ are assumed known for $T_2$ and vice versa for the estimation of $\lambda_2$ and $\alpha_2$. Noninformative  censoring times were generated from a uniform distribution $(0,\gamma)$, where $\gamma$ was chosen for $10,\ 30 $ (omitted from table) and $ 50\%$ censoring. We consider the Gumbel copula with Kendall's tau, $\tau_{ken} = 1-\frac{1}{\theta} =  0, \ 0.25, \ 0.50$, and $\ 0.75 $. Initial parameter values were randomly chosen from uniform distributions, with multiple starting values (wherever possible)  as described in Section 3. We also simulated data from the Clayton copula.  The results are similar to those for the Gumbel copula and are described in the appendix. 
Tables \ref{t:1}, \ref{t:5}  and \ref{t:6} show the results for estimation of the model for $T_1$ 
treating $T_2$ as a competing event and for $T_2$ treating $T_1$ (for brevity, tables  \ref{t:5}  and \ref{t:6} show results for selected sample sizes)  as a competing event. The bias is small decreasing to zero as the sample size increases for each of the censoring levels. The empirical variance and the model based variance tend to agree and the coverage is close to the nominal $0.95$ level, particularly at larger sample sizes. The empirical variance decreases as the sample size increases at roughly the expected root n rate.  Table 4 shows the robust survival estimates for the mis-specified model (data simulation from exponentiated Weibull but Weibull survival estimated instead) for $T_1$ treating $T_2$ as competing risk.  We presented the results for the lower, mean and upper quantiles for each of the sample sizes and
dependence levels.  This shows decreasing bias for increasing sample sizes for each of the quantiles.

\begin{table}
\caption{\label{t:1} Estimated parameters for the exponentiated Weibull model for $T_1$ across samples sizes (N), dependence levels ($\tau_{ken}$) and  for 15\%  censoring treating $T_2$ as a competing event.  $\hat{\eta}$: estimated parameters: $\lambda$, scale,$\alpha$, and $\kappa$  are shape parameters, ModB: model-based variance, EMP: 
empirical variance, CP: $95\%$ coverage probability.} 
\centering
\begin{tabular}{cccccccc} \hline
$\tau_{ken} $ & N &$\hat{\eta}$& Mean & Bias$$ & ModB$$ & EMP $$ & CP 
\\ \hline 
 
0.00 & 1000 & $\hat{\lambda}$ & 2.936 & -0.064 & 0.069 & 0.056 & 0.960  \\
& & $\hat{\kappa}$  & 4.198 &  0.198 & 8.663 & 3.873 & 0.848  \\
& &  $\hat{\alpha}$  & 0.120 &  0.020 & 0.004 & 0.003 & 0.942  \\
& 2500 & $\hat{\lambda}$ & 2.965 & -0.035 & 0.025 & 0.023 & 0.949  \\
&&  $\hat{\kappa}$  & 4.180 &  0.180 & 2.839 & 2.305 & 0.885  \\
& & $\hat{\alpha}$  & 0.109 &  0.009 & 0.001 & 0.002 & 0.894  \\
& 5000 & $\hat{\lambda}$ & 2.994 & -0.006 & 0.012 & 0.012 & 0.947  \\
& & $\hat{\kappa}$  & 4.264 &  0.264 & 1.446 & 1.250 & 0.947  \\
& & $\hat{\alpha}$  & 0.100 &  1.6e-4 & 0.001 & 0.001 & 0.931  \\ \\

0.25&1000 & $\hat{\lambda}$ & 3.001 &  0.001 & 0.051 &  0.062 & 0.924  \\
 & & $\hat{\kappa}$  & 4.912 &  0.912 & 3.042 & 14.274 & 0.922  \\
 & & $\hat{\alpha}$  & 0.101 &  0.001 & 0.002 &  0.002 & 0.896  \\
& 2500 & $\hat{\lambda}$ & 3.005 &  0.005 & 0.019 &  0.018 & 0.960  \\
 & & $\hat{\kappa}$  & 4.213 &  0.213 & 0.765 &  0.829 & 0.958  \\
& & $\hat{\alpha}$  & 0.100 &  0.000 & 0.000 &  0.001 & 0.928  \\
& 5000 & $\hat{\lambda}$ & 2.998 & -0.002 & 0.009 &  0.008 & 0.960  \\
& &  $\hat{\kappa}$  & 4.082 &  0.082 & 0.300 &  0.297 & 0.958  \\
& & $\hat{\alpha}$  & 0.100 &  1.0e-5 & 2.0e-4 &  2.5e-4 & 0.958  \\\\

0.50 & 1000 & $\hat{\lambda}$ & 3.003 &  0.003 & 0.036 & 0.038 & 0.954  \\
& &  $\hat{\kappa}$  & 4.376 &  0.376 & 1.645 & 1.898 & 0.952  \\
& & $\hat{\alpha}$  & 0.100 &  3.3e-4 & 0.001 & 0.001 & 0.920  \\
& 2500 & $\hat{\lambda}$ & 3.006 &  0.006 & 0.014 & 0.014 & 0.950  \\
& & $\hat{\kappa}$  & 4.147 &  0.147 & 0.439 & 0.523 & 0.938  \\
& & $\hat{\alpha}$  & 0.100 &  1.7e-5 & 2.6e-4 & 0.001 & 0.934  \\
& 5000  & $\hat{\lambda}$ & 3.002 &  0.002 & 0.007 & 0.007 & 0.964  \\
& & $\hat{\kappa}$  & 4.072 &  0.072 & 0.185 & 0.187 & 0.948  \\
& & $\hat{\alpha}$  & 0.099 & -5.7e-4 & 1.2e-4 & 1.4e-4 & 0.940  \\\\

0.75 & 1000 & $\hat{\lambda}$ & 3.003 &  0.003 & 0.027 & 0.042 & 0.942  \\
& & $\hat{\kappa}$  & 4.266 &  0.266 & 1.101 & 1.146 & 0.950  \\
& & $\hat{\alpha}$  & 0.102 &  0.002 & 0.001 & 0.007 & 0.926  \\
& 2500 & $\hat{\lambda}$ & 3.006 &  0.006 & 0.011 & 0.010 & 0.942  \\
& & $\hat{\kappa}$  & 4.132 &  0.132 & 0.323 & 0.330 & 0.954  \\
& & $\hat{\alpha}$  & 0.099 & -0.001 & 2.0e-4 & 2.0e-4 & 0.938  \\
& 5000 & $\hat{\lambda}$ & 3.004 &  0.004 & 0.005 & 0.005 & 0.952  \\
& & $\hat{\kappa}$  & 4.066 &  0.066 & 0.145 & 0.145 & 0.956  \\
& & $\hat{\alpha}$  & 0.099 & -0.001 & 9.7e-5 & 1.1e-4 & 0.934  \\  \hline 
\end{tabular}
\end{table}

\subsection{Application to French Breast Cancer Data}
In this section we analyze data from women between the ages of $18$ and $96$ years surviving breast cancer in France 
from 1980 to 2011. The data were obtained from the Institut Curie breast cancer database. This database contains records from $24,458$ nonmetastatic breast cancer patients treated  at the Institut Curie. Out of the $24,458$ breast cancer patients, $9,885 \ (40.4\%)$ died  while $14,573$ were alive and administratively censored on December $31^{st} \ 2011$. Five age group categories were considered for the estimation of relative survival.  $3,970$ were between the ages of $15-44$, $6,895$ between the ages of $45-54$, $6,420$ between the ages of $55-64$, $4,675$ between the ages of $65-74$ and $2,498$ were in the $75-99$ age group category. We individually matched the observed death or censoring time in the disease cohort group with a corresponding time in the general reference population on age, sex, and year (date of diagnosis and the date of death or censored) for each participant and for each follow-up period. The background mortality data from the Human Mortality Database (https://www.mortality.org) was last modified on June 28, 2018.  Within each follow-up year, we assumed that $\lambda_P(t)$ is piecewise constant (Dickman {\it et al.}, 2004) 
for each period up to time X. The cumulative hazard for each period based on $\lambda_P(t)$ is calculated from the background survival function at the beginning and end of the period. The cumulative hazard is then used to obtain $\lambda_P(t)$ under the piecewise constant assumption. The goal of matching in determining $\lambda_{T_2}=\lambda_P$ is to mitigate the impact of age and calendar year on potentially dependent censoring by C (Pohar Perme et al., 2012). 
Thus, our approach is stratified by age, sex and year. We estimate $2, 5, 10$, and $15-$year relative survival assuming a Weibull distribution for $T_1$ and a Gumbel copula model with differing levels of dependence to specify the joint distribution for  $T_1$ and $T_2$. We compared our estimates  with estimates from  Pohar Perme et al. (2012),  
which require independence of $T_1$ and $T_2$ with  $S_{T_2}(t)$  derived from the background reference population.

Tables \ref{t:2} and \ref{t:3} show the estimates of $S_{T_1}(t)$ for cancer mortality both overall and stratified by age. The parametric estimates under independence are similar to those from the Pohar-Perme method. 
This suggests that the Weibull assumption is a reasonable fit to the data. One observes that as dependence increases, cancer survival generally decreases. For a fixed dependence level, younger women tend to have higher cancer survival rates than do older women, with marked reductions for the 65-74 and 75-99 age groups. There is some instability in survival estimates at 15 years, especially for the older age groups, as evidenced by the large standard errors. This may be due to small numbers of patients at risk at longer follow-up times.

\begin{table}
\caption{\label{t:2} The 2, 5, 10 and 15-yr overall net survival for French women diagnosed with breast cancer between 1980 and 2011. $a: \times 10^{-2}$, $b: \times 10^{-3}$, $\tau_{ken}$: dependence, PP: Pohar-Perme, $S_{T_1}(t)$: survival estimate at year t, SE: standard error for the relative survival estimate.}
\centering
\begin{tabular}{rrrrrrrrrr}
\hline
&\multicolumn{3}{c}{Independent Competing Risks} &
\multicolumn{6}{c}{Dependent Competing Risks}\\ 
$\tau_{ken}$ & \multicolumn{3}{c}{$0.00$ } &
\multicolumn{2}{c}{$0.25$ } &
\multicolumn{2}{c}{$0.50$ } &
\multicolumn{2}{c}{$0.75$ }\\
\hline
t & $PP^a$ & $S_{T_1}(t)^a$ & $SE^b$ & $S_{T_1}(t)^a$ & $SE^b$ & $S_{T_1}(t)^a$ & $SE^b$ & $S_{T_1}(t)^a $& $SE^b$  \\  \hline
2 & 95.6  & 96.0 & 6.99 & 95.8 & 6.96 & 95.4 & 7.23 & 94.7 & 7.74 \\
5 & 84.8  & 87.4 & 9.01 &  86.6 & 9.10 &  85.5 & 9.31 & 84.0 & 9.53 \\
10 & 71.0 & 72.8 & 11.01 & 71.4 & 10.99 &  69.8 & 10.91 & 68.0 & 10.67 \\
15 & 59.5 & 59.5 & 12.22 &  57.9 &  12.08 & 56.3 & 11.74 & 54.9 & 11.19 \\
\hline 
\end{tabular} 
\end{table}

\begin{table}
\caption{ \label{t:3} The 2, 5, 10 and 15-yr  age group specific net breast cancer survival among French women diagnosed between 1980 and 2011.  $a:  \times 10^{-2}$, $b: \times 10^{-3}$, $\tau_{ken}$: dependence, PP: Pohar-Perme, $S_{T_1}(t)$: relative survival estimate at year t,  SE: standard error for the relative survival estimate.}
\centering
\begin{tabular}{lrrrrrrrrrr}
\hline 
& & \multicolumn{3}{c}{Independent Competing  Risks} &
\multicolumn{6}{c}{Dependent Competing Risks }\\  
$\tau_{ken} $ & &\multicolumn{3}{c}{$0.00$ } & 
\multicolumn{2}{c}{$0.25$ } & 
\multicolumn{2}{c}{$0.50$ } & 
\multicolumn{2}{c}{$0.75$ } \\ \hline 
t & Agegp & $PP^a$  & $S_{T_1}(t)^a$ & $SE^b$ & $S_{T_1}(t)^a$ & $SE^b$ & $S_{T_1}(t)^a$ & $SE^b$ & $S_{T_1}(t)^a $& $SE^b$  \\  \hline 
2 & 15-44 & 95.8  & 94.9 & 20.90 & 94.9 & 20.73 & 94.8 & 20.73 & 94.8 & 20.68 \\
& 45-54 & 97.1  & 96.6 & 16.44 & 96.5 & 16.13 & 96.3 & 16.27 & 96.2 & 16.40 \\
& 55-64 & 95.7 & 96.1 & 13.72 & 96.0 & 13.49 & 95.7 & 13.70 & 95.3 & 14.12 \\
& 65-74 & 95.1 & 97.0 & 08.50 & 96.8 & 08.54 & 96.2 & 09.61 & 95.1 & 11.60 \\
& 75-99 & 91.5 & 96.5 & 07.94 & 95.6 & 08.93 & 93.4 & 12.44 & 89.9 & 17.16 \\ \\

5 & 15-44 & 85.1 & 86.9 & 23.70 & 86.8 & 23.64 & 86.7 & 23.62 & 86.7 & 23.35 \\
& 45-54 & 88.6  & 90.4 & 19.39 & 90.1 & 19.36 & 89.8 & 19.45 & 89.7 & 19.28 \\
& 55-64 & 85.8  & 88.1 & 17.72 & 87.6 & 17.71 & 86.9 & 17.87 & 86.6 & 17.66 \\
& 65-74 & 84.1  & 86.9 & 16.71 & 85.8 & 17.01 & 84.2 & 17.71 & 82.5 & 18.09 \\
& 75-99 & 72.3  & 77.1 & 24.21 & 72.7 & 24.85 & 67.1 & 25.08 & 61.7 & 24.00 \\ \\

10 & 15-44 & 71.9 & 74.4 & 26.88 & 74.2 & 26.84 & 74.0 & 26.75 & 74.1 & 26.62 \\
& 45-54 & 78.3  & 80.1 & 22.83 & 79.6 & 22.80 & 79.2 & 22.73 & 79.2 & 22.34 \\
& 55-64 & 73.4  & 74.5 & 22.03 & 73.5 & 21.97 & 72.7 & 21.74 & 72.7 & 21.08 \\
& 65-74 & 68.4  & 67.2 & 25.38 & 65.0 & 25.32 & 63.0 & 24.72 & 62.3 & 23.20 \\
& 75-99 & 44.6  & 43.1 & 34.83 & 37.0 & 32.55 & 33.0 & 28.61 & 31.1 & 24.35 \\ \\

15 & 15-44 & 62.5 & 63.2 & 29.03 & 63.0 & 28.96 & 62.9 & 28.83 & 63.0 & 28.72 \\
& 45-54 & 70.8  & 70.5 & 25.31 & 69.8 & 25.24 & 69.4 & 25.00 & 69.6 & 24.51 \\
& 55-64 & 63.5  & 61.9 & 24.81 & 60.7 & 24.62 & 59.9 & 24.11 & 60.3 & 23.20  \\
& 65-74 & 50.3  & 48.7 & 30.06 & 46.2 & 29.46 & 44.7 & 27.92 & 45.3 & 25.47 \\
\hline 
\end{tabular}
\end{table}

\begin{table}
\caption{\label{t:4} Robust survival probability for a misspecified exponentiated Weibull model for $T_1$ across samples sizes (N), dependence levels ($\tau_{ken}$) and  for 16\%  censoring treating $T_2$ as a competing event.  $\hat{S_w } $: estimated  Weibull survival probability, and associated bias at the  $0.25, 0.50$, and $0.75$ quantiles.}
\centering
\begin{tabular}{cccccccccc}  \hline
& &  \multicolumn{3}{c}{Survival Quantiles} &
\multicolumn{3}{c}{Bias Quantiles }\\  
$\tau_{ken} $ &   N &   
   $\widehat{ sw.25 }$ &   $\widehat{sw.50 }$ &  $\widehat{sw.75 }$  & Bias.25  &Bias.50 & Bias.75
\\ \hline 
 
0.00 &  1000 
& 0.373 & 0.532 & 0.751 & -19.5e-4 & 25.4e-3 & 11.7e--3  \\
 &  2500 
 & 0.372 & 0.532 & 0.750 & -16.8e-4 & 26.0e-3 & 12.3e-3  \\
 &  5000 
 & 0.372 & 0.532 & 0.751 & -15,1e-4 & 26.1e-3 & 12.3e-3  \\ \\

0.25 &  1000 
 & 0.363 & 0.533 & 0.755 & -14.7e-3 & 18.5e-3 & 68.6e-4  \\
 &   2500 
  & 0.362 & 0.532 & 0.754 & -14.2e-3 & 19.1e-3 & 74.9e-4  \\
 & 5000
  & 0.362 & 0.532 & 0.754 & -13.6e-3 & 19.6e-3 & 78.0e-4  \\\\

0.50 &  1000
 & 0.345 & 0.527 & 0.758 & -17.1e-3 & 17.8e-3 & 31.2e-4  \\
&  2500
 & 0.345 & 0.527 & 0.758 & -16.7e-3 & 18.4e-3 & 36.2e-4  \\
&  5000
 & 0.345 & 0.527 & 0.758 & -16.3e-3 & 18.7e-3& 38.7e-4  \\ \\

0.75 & 1000
 & 0.326 & 0.521 & 0.761 & -74.6e-4 & 22.5e-3 & -5.1e-4  \\
&  2500
  & 0.325 & 0.519 & 0.761 & -64.8e-4 & 23.4e-3 & -1.9e-05  \\
& 5000
 & 0.325 & 0.519 & 0.761 & -61.6e-4 & 23.8e-3 & 2.1e-05  \\
 \hline 
\end{tabular}
\end{table}

The net survival function for cancer corresponds to a hypothetical world where the only cause of death is breast cancer. This quantity can only be estimated under unverifiable dependence assumptions between $T_1$ and $T_2$ using disease registry data. To account for uncertainty in dependence, we recommend reporting a range of probabilities corresponding to differing levels of dependence. For example, using results from table $2$, the overall $5$ year net breast cancer survival from $1980 - 2011$ is estimated to be between 84.0-87.4\% under dependence ranging from Kendall's tau equal to $0$ (independence) to $0.75$ (strong dependence). These cancer survival probabilities may be meaningfully compared with those in other populations having different background mortality rates and different dependence levels between $T_1$ and $T_2$.

The sensitivity analysis was conducted across different levels of dependence representing different competing mortality potentially observable in the registry data. Figures $2$  and $3$ show the $2, \ 5, \ 10$ and $15$-yr overall net breast survival plots across a spectrum of dependence structures for women between the ages of 18 and 96-yr living in France during 2008 and 2011. As the dependence level increases, the net breast cancer survival decreases dramatically. Perhaps, this might be due to increase hazard for the patients compounding the effect of competing mortality hereby decreasing the chances of survival.  

\section{Discussion and Conclusion}
\label{s:discuss}
Our model formulation for competing risk data without cause of failure information is general, permitting arbitrary but known copula functions. The distribution of other cause mortality is obtained from external reference data 

(Sarfati {\it et al.}, 2010; Pohar Perme {\it et al.}, 2012; Sasieni and Brentnall, 2017).
We have undertaken preliminary investigations of simultaneous estimation of the dependence parameter and the parameter in the disease-specific survival distribution. There is evidence of instability in the estimation process especially at the boundary values, with care needed in the model specification to aid estimability of the model parameters. This is expected (see Zahl, 1997 for challenges), as there are similar issues even when the cause of failure is known. The proposed sensitivity analysis is a practical solution to this issue, providing a range of estimates across different dependence levels not requiring simultaneous estimation of the dependence parameter. The parametric model for disease-specific mortality is restrictive but may be flexible enough for applications where the hazard is smooth over time, which is the case in cancer registry data.  Estimating expected hazard or the distribution of $T_2$ from life tables is limiting as mismatches in covariates and other stratifying variables may also induce biases, Rubio et al (2021). However, it is important to note that our estimator is robust and arguably invariant to the parameter estimates under different specifications of the copula models as shown in the parameter estimates for both Gumbel and Clayton copulas (tables 5 and 6). To relax the parametric assumption, nonparametric techniques are currently being developed for use in more complex failure patterns.

The focus of relative or net survival analysis is the distribution of the latent event time for death from disease. This endpoint has been advocated by many practitioners 
(Slud {\it et al.}, 1988; Reason, 1990; and Louzada {\it et al.}, 2015), as it removes the impact of other cause mortality on the risk of disease-specific mortality, permitting comparisons across populations with different background mortality. As an alternative, other work has considered estimation of the crude disease-specific survival, $C_k(t)$, using the relative survival estimates and the known reference hazard for other cause mortality (Cronin and Feuer, 2000).  
An analogous procedure could be implemented using our copula based estimate of the distribution of $T_1$ and would provide an assessment of the sensitivity of the estimator of $C_k$ under independence of $T_1$ and $T_2$. Such procedure would be of interest to individuals who prefer crude disease-specific mortality to net disease-specific mortality. This and time-dependent models are  topics for future research. 

Our proposed estimator performed well overall and by subgroup analysis. We observed that in cases of elderly patients, long-term survival decreases dramatically as expected. Perhaps this might be due to elderly patients experiencing higher expected mortality rates than younger patients particularly in terms of higher risk of death from other competing causes leading to loss of patients at longer follow-up times. This loss of information induces higher variability in net survival estimates for this elderly populations, thereby inducing a higher variability as observed in the variance. Net survival under dependence competing risk assumption is observable and does not require additive model as in the case for net survival (in hypothetical world) under the independence assumption. Our estimator modelled both the dependence between times to disease-specific event and competing risk event with covariates like age, sex, and period (date of diagnosis and date of event: death or censored). Since it’s been known (Nanieli, et. al., 2012) that such covariates affect both excess and expected hazards estimates. We conditioned on these covariates (Pohar-Perme (2012) by matching cancer cohort data with registry data to alleviate biases associated with informative censoring as done in multivariable modelling techniques (Bolard, et. al., (2002),  Giorgi, et. al., (2003),  Lambert, et. al., (2005),  Remontet, et. al., (2007))  where estimators are adjusted for life-table covariates.  Researchers are encouraged to use a patients' medical history in determining the levels of competing risks and use a corresponding dependence survival estimate (independence, low, moderate or high) as a measure for disease-specific prognosis.

In conclusion, our proposed methodology provides estimates for net survival under both independent and dependent  competing  mortality. On the contrary, Pohar-Perme et al., (2012) estimator is only valid under the independent competing risk assumption. Additionally, Pohar-Perme et al., (2012) estimator may exceed 1 in the left tail.  Schaffar  et al., (2017) showed that these erratic results may occur with longer follow-up times. Our estimator provides comparable  relative  or net survival estimates under both independent and dependent competing risk assumptions  without the need for cause of disease-specific event in the competing risks  setting.

\vspace*{1pc}

\noindent {\bf{Conflict of Interest}}

\noindent {\it{ None declared }}

\section*{Appendix}
\begin{table}
\caption{\label{t:5} Gumbel Model: Estimated parameters of the Weibull model for $T_1$ across samples sizes (N), dependence levels ($\tau_{ken}$) and levels of censoring (C) treating $T_2$ as a competing event and vice versa.  $\hat{\eta}$: estimated parameters, ModB: model-based variance, EMP: empirical variance, CP: $95\%$ coverage probability. $^a:\times 10^{-3}$.} 
\centering
\begin{tabular}{ccccccccccccc} \hline
& C & & \multicolumn{5}{c}{$0.10$ } &
\multicolumn{5}{c}{$0.50$ } \\ \hline
$\tau_{ken} $ & N &$\hat{\eta}$& Mean & Bias$^a$ & ModB$^a$ & EMP $^a$ & CP 
& Mean & Bias${^a}$ & ModB${^a}$ & EMP${^a}$& CP \\ \hline 

0.00 & 1000 & $\hat{\lambda_1}$ & 0.182 & -0.080 & 0.090 & 0.090 & 0.940  
& 0.182 & -0.290 & 0.150 & 0.170 & 0.928  \\
& & $\hat{\alpha_1}$ & 1.610 &  0.790 & 1.420 & 1.560 & 0.938  
& 1.611 &  1.980 & 2.620 & 2.720 & 0.950  \\

& 5000 & $\hat{\lambda_1}$ & 0.182 & -0.050 & 0.020 & 0.020 & 0.948  
& 0.182 & 0.050 & 0.030 & 0.030 & 0.958  \\
& & $\hat{\alpha_1}$ & 1.610 &  0.520 & 0.280  & 0.280 & 0.954  
& 1.610 & 0.290 & 0.520 & 0.530 & 0.956  \\ 
\hline

& 1000  & $\hat{\lambda_2}$ & 0.748 & 5.980 & 9.940 & 10.270 & 0.936 
& 0.746 & 3.650 & 14.410 & 15.320 & 0.922  \\
& & $\hat{\alpha_2}$ & 0.694 & 0.840 & 6.790 & 7.020 & 0.944   
& 0.697 & 4.070 & 8.490 & 0.010 & 0.948  \\

& 5000 & $\hat{\lambda_2}$ & 0.743 & 0.630 & 1.870 & 1.640 & 0.962 
& 0.743 & 1.000 & 2.700 & 2.460 & 0.968  \\ 
& &$\hat{\alpha_2}$ & 0.693 & 0.100 & 1.340 & 1.210 & 0.962  
& 0.693 & 0.280 & 1.680 & 1.530 & 0.958  \\  \\

0.25 & 1000 & $\hat{\lambda_1}$ & 0.182 & -0.480 & 0.080 & 0.080 & 0.948  
& 0.182 & -0.450 & 0.140 & 0.130 & 0.954  \\
& & $\hat{\alpha_1}$ & 1.610 &  1.490 & 1.310 & 1.340 & 0.952 
& 1.613 &  3.960 & 2.480 & 2.840 & 0.934  \\

& 5000 & $\hat{\lambda_1}$ & 0.182 & -0.050 & 0.020 & 0.020 & 0.956 
& 0.182 & -0.130 & 0.030 & 0.030 & 0.940  \\
& & $\hat{\alpha_1}$ & 1.609 & -0.250 & 0.260 & 0.240 & 0.956 
& 1.610 & 0.590 & 0.500 & 0.460 & 0.950  \\ 
\hline 

& 1000 & $\hat{\lambda_2}$ & 0.753 &  10.920 & 14.700 & 14.430 & 0.938   
& 0.760 &  18.430 & 20.460 & 20.810 & 0.946  \\
&  & $\hat{\alpha_2}$ & 0.690& -3.170 & 8.240 & 7.930 & 0.954 
& 0.687 & -5.930 & 10.080 & 9.900 & 0.944  \\

& 5000 & $\hat{\lambda_2}$& 0.741 & -0.950 & 2.690 & 2.530 & 0.950  
& 0.742 & 0.260 & 3.620 & 3.580 & 0.964  \\
& & $\hat{\alpha_2}$ & 0.695 &  1.940 & 1.610 & 1.480 & 0.958 
& 0.695 & 2.260 & 1.960 & 1.940 & 0.948   \\  \\

0.50 & 1000 & $\hat{\lambda_1}$ & 0.182 & -0.030 & 0.070 & 0.070 & 0.956 
& 0.182 & -0.260 & 0.120 & 0.120 & 0.954  \\
& & $\hat{\alpha_1}$ & 1.611 &  2.270 & 1.270 & 1.300 & 0.948 
& 1.613 &  3.840 & 2.380 & 2.710 & 0.928  \\

& 5000 & $\hat{\lambda_1}$ & 0.182 &  0.010 & 0.010 & 0.020 & 0.946 
& 1.824 & 0.040 & 0.020 & 0.030 & 0.956  \\
& & $\hat{\alpha_1}$ & 1.609 & -0.340 & 0.250 & 0.240 & 0.954  
& 1.610 & 0.450 & 0.480 & 0.510 & 0.932  \\ 
\hline

& 1000 & $\hat{\lambda_2}$ & 0.759 &  17.440 & 19.080 & 20.140 & 0.932 
& 0.767 &  25.540 & 25.050 & 25.330 & 0.932  \\
& & $\hat{\alpha_2}$  & 0.688 & -5.180 & 9.440 & 9.910 & 0.944  
& 0.684 & -9.170 & 11.210 & 11.510 & 0.940  \\

& 5000 & $\hat{\lambda_2}$& 0.740 & -1.580 & 3.360 & 3.380 & 0.944 
& 0.744 &  1.620 & 4.340 & 4.660 & 0.946  \\
& & $\hat{\alpha_2}$ & 0.695 &  1.720 & 1.820 & 1.870 & 0.936   
& 0.692 & -0.750 & 2.150 & 2.270 & 0.944  \\ \\


0.75 & 1000 & $\hat{\lambda_1}$  & 0.182 & -0.200 & 0.060 & 0.070 & 0.956 
& 0.182 & -0.260 & 0.100 & 0.100 & 0.948  \\
& & $\hat{\alpha_1}$ & 1.610 &  0.490 & 1.060 & 1.520 & 0.936 
& 1.612 &  2.660 & 2.090 & 2.370 & 0.942  \\

& 5000 & $\hat{\lambda_1}$  & 0.182 &  0.050 & 0.010 & 0.010 & 0.948  
& 0.182 & -0.020 & 0.020 & 0.020 & 0.952  \\
& & $\hat{\alpha_1}$ & 1.609& -0.010 & 0.210 & 0.210 & 0.944 
& 1.609 & -0.120 & 0.420 & 0.450 & 0.948  \\ 

\hline 

& 1000 & $\hat{\lambda_2}$  & 0.760 &  17.780 & 20.190 & 20.360 & 0.938  
& 0.766 &  24.200 & 26.040 & 25.790 & 0.952  \\
& & $\hat{\alpha_2}$ & 0.689 & -4.600 & 9.870 & 10.440 & 0.946  
& 0.685 & -7.510 & 11.580 & 11.590 & 0.956  \\

& 5000 & $\hat{\lambda_2}$  & 0.742 & -0.190 & 3.540 & 3.770 & 0.946 
& 0.743 & 0.830 & 4.440 & 4.540 & 0.950  \\
&  & $\hat{\alpha_2}$ & 0.694 &  0.550 & 1.900 & 1.990 & 0.930  
& 0.693 & 0.350 & 2.210 & 2.280 & 0.944  \\  \hline 
\end{tabular}

\end{table}

\subsection*{Simulation Results for  Gumbel and Clayton Copula Models}

We simulated data to mimic the French breast cancer data set for sample sizes; $1000$,  and $5000$ with $500$ replications. The latent failure times for $T_j \sim Weibull(\alpha_j, \lambda_j)$ with probability density function defined in section $3$. The parameters for the Weibull distribution for $T_1$ were $ \lambda_1 = 0.182$ and  $\alpha_1 = 1.609$, while those for  $T_2$ were $\lambda_2 = 0.742 $ and $\alpha_2 = 0.693 $. In the estimation of $\lambda_1, \ \alpha_1$ for $T_1$, $\lambda_2, \ \alpha_2$ are assumed known for $T_2$, and vice versa for estimation of $\lambda_2$ ,$\alpha_2$. Noninformative censoring times were generated from a uniform distribution $(0,\gamma)$, where $\gamma$ was chosen for $10,\ 30 $ and $ 50\%$ censoring. 
We consider the Clayton copula with Kendall's tau, $\tau_{ken} = \frac{\theta}{\theta + 2} =  0,\ 0.25,\ 0.50,\ 0.75$. Initial parameter values were randomly chosen from uniform distributions, with multiple starting values  as described in section 3. The simulation results based on the Clayton copula are presented in the table \ref{t:4} below.

\begin{table}
\caption{\label{t:6} Clayton Model: Estimated parameters of the Weibull model for $T_1$ across samples sizes (N), dependence levels ($\tau_{ken}$) and levels of censoring (C) treating $T_2$ as a competing event and vice versa.  $\hat{\eta}$: estimated parameters, ModB: model-based variance, EMP: empirical variance, CP: $95\%$ coverage probability. $^a:\times 10^{-3}$.} 
\centering
\begin{tabular}{lcccccccccccc} \hline
& C & & \multicolumn{5}{c}{$0.10$ } &
\multicolumn{5}{c}{$0.50$ } \\ \hline
$\tau_{ken} $ & N &$\hat{\eta}$& Mean & Bias$^a$ & ModB$^a$ & EMP $^a$ & CP 
& Mean & Bias${^a}$ & ModB${^a}$ & EMP${^a}$& CP \\ \hline 

0.00 & 1000 & $\hat{\lambda_1}$ & 0.182 & 0.000 & 0.080 & 0.090 & 0.948 
& 0.182 & -0.340 & 0.140 & 0.160 & 0.930 \\

& & $\hat{\alpha_1}$ & 1.610 & 0.710 & 1.390 & 1.520 & 0.942 
& 1.611 &  1.820 & 2.490 & 2.540 & 0.948   \\ 

& 5000 & $\hat{\lambda_1}$ & 0.182 & -0.020 & 0.020 & 0.020 & 0.942 
& 0.182 & 0.060 & 0.030 & 0.030 & 0.946 \\

& & $\hat{\alpha_1}$ & 1.610 &  0.520 & 0.280 & 0.280 & 0.956 
& 1.610 & 0.710 & 0.500 & 0.510 & 0.952  \\ 
\hline 
& 1000 &$\hat{\lambda_1}$ & 0.747 & 5.910 & 9.750 & 9.910 & 0.940  & 0.746 & 3.570 & 14.010 & 14.150 & 0.922  \\
& &  $\hat{\alpha_1}$ & 0.694 & 0.780 & 6.720 & 6.950 & 0.948   & 0.696 & 3.840 & 8.360 & 8.590 & 0.956  \\ 

& 5000 & $\hat{\lambda_1}$ & 0.742 & 0.690 & 1.840 & 1.610 & 0.962 & 0.742 & 0.880 & 2.640 & 2.380 & 0.964  \\
& & $\hat{\alpha_1}$ & 0.693 & 0.050 & 1.330 & 1.200 & 0.962  & 0.693 & 0.320 & 1.650 & 1.490 & 0.956  \\ \\
0.25 & 1000 & $\hat{\lambda_1}$  & 0.182 & -0.010 & 0.070 & 0.080 & 0.952 
& 0.182 & -0.670 & 0.130 & 0.110 & 0.964 \\

& & $\hat{\alpha_1}$  & 1.610 &  0.560 & 1.280 & 1.410 & 0.930 
& 1.611 &  1.870 & 2.390 & 0.220 & 0.952   \\ 

& 5000 & $\hat{\lambda_1}$ & 0.182 & -0.180 & 0.010 & 0.010 & 0.946 
& 0.182 & -0.120 & 0.020 & 0.020 & 0.940\\

& & $\hat{\alpha_1}$ & 1.609 & -0.310 & 0.230 & 0.230 & 0.950  
& 1.609 & -0.030 & 0.430 & 0.450 & 0.958 \\ 
\hline 

& 1000 & $\hat{\lambda_1}$ & 0.751 & 9.200 & 16.630 & 17.220 & 0.924 & 0.749 & 7.170 & 21.000 & 1.610 & 0.914  \\
& & $\hat{\alpha_2}$ & 0.693 & 0.200 & 8.520 & 8.840 & 0.940  & 0.696 & 2.890 & 10.090 & 10.680 & 0.942  \\

& 5000 & $\hat{\lambda_2}$ & 0.742& 0.460 & 2.980 & 2.690 & 0.952 & 0.744 & 2.080 & 3.840 & 3.690 & 0.940  \\
& & $\hat{\alpha_2}$ & 0.693 & 0.550 & 1.660 & 1.54 & 0.950 & 0.693 & 0.140 & 1.980 & 1.930 & 0.942  \\  \\

0.50 & 1000 &  $\hat{\lambda_1}$ & 0.182 & -0.090 & 0.060 & 0.050 & 0.956 
& 0.182 & -0.440 & 0.100 & 0.100 & 0.964 \\

& & $\hat{\alpha_1}$ & 1.608 & -0.980 & 1.030 & 1.090 & 0.952 
& 1.612 &  2.580 & 2.020 & 2.060 & 0.968  \\

& 5000 & $\hat{\lambda_1}$ & 0.182 & -0.180 & 0.010 & 0.010 & 0.936 
& 0.182 & -0.180 & 0.020 & 0.020 & 0.940 \\

& & $\hat{\alpha_1}$ & 1.609 & -0.190 & 0.200 & 0.210 & 0.952 
& 1.610 &  0.240 & 0.400 & 0.410 & 0.948 \\
\hline

& 1000 & $\hat{\lambda_2}$ & 0.750 & 9.020 & 17.470 & 18.370 & 0.924 & 0.748 & 6.530 & 21.500 & 22.090 & 0.912  \\
& & $\hat{\alpha_2 }$ & 0.693 & 0.390 & 8.850 & 9.390 & 0.928 & 0.696 & 3.120 & 10.350 & 11.080 & 0.930  \\

& 5000 & $\lambda_2$ & 0.742 & 0.070 & 3.170 & 2.870 & 0.948  & 0.744 &  2.400 & 3.970 & 3.760 & 0.944  \\
& & $\alpha_2$ & 0.693 & 0.460 & 1.730 & 1.620 & 0.950  & 0.693 & -0.080 & 2.040 & 1.960 & 0.948  \\  \\

0.75 & 1000 & $\hat{\lambda_1}$   & 0.182 & 0.130 & 0.040 & 0.040 & 0.944  
& 0.182 & -0.050 & 0.080 & 0.080 & 0.937 \\

& & $\hat{\alpha_1}$ & 1.609 & 0.030 & 7e-04 & 0.860 & 0.924  

& 1.610 &  1.120 & 1.410 & 1.450 & 0.947 \\

& 5000 & $\hat{\lambda_1}$  & 0.182 & -0.120 & 0.010 & 0.010 & 0.940 
& 0.182 & -0.040 & 0.020 & 0.020 & 0.948 \\

& & $\hat{\alpha_1}$ & 1.609 &  0.040 & 0.140 & 0.160 & 0.936 

& 1.610 &  0.780 & 0.270 & 0.320 & 0.926 \\  

\hline
& 1000 &  $\hat{\lambda_1}$ & 0.747 & 4.900 & 13.000 & 14.790 & 0.924 & 0.744 & 1.780 & 16.030 & 17.270 & 0.928  \\
& & $\hat{\alpha_1}$ & 0.694 & 1.460 & 8.370 & 9.630 & 0.924  & 0.697 & 4.230 & 9.660 & 10.980 & 0.932 \\ 

& 5000 & $\hat{\lambda_1}$  & 0.743 &  1.160 & 2.440 & 2.120 & 0.966 & 0.744 &  1.920 & 3.040 & 2.590 & 0.966  \\
&& $\hat{\alpha_1}$ & 0.693 & -0.090 & 1.650 & 1.490 & 0.970  & 0.693 & -0.470 & 1.910 & 1,700 & 0.962  
\\
\hline 
\end{tabular}

\end{table}

\bigskip
 
\newpage

\begin{figure}[ht]
\label{Figure:1}
\centering
\includegraphics[width = 15cm, height = 15cm]{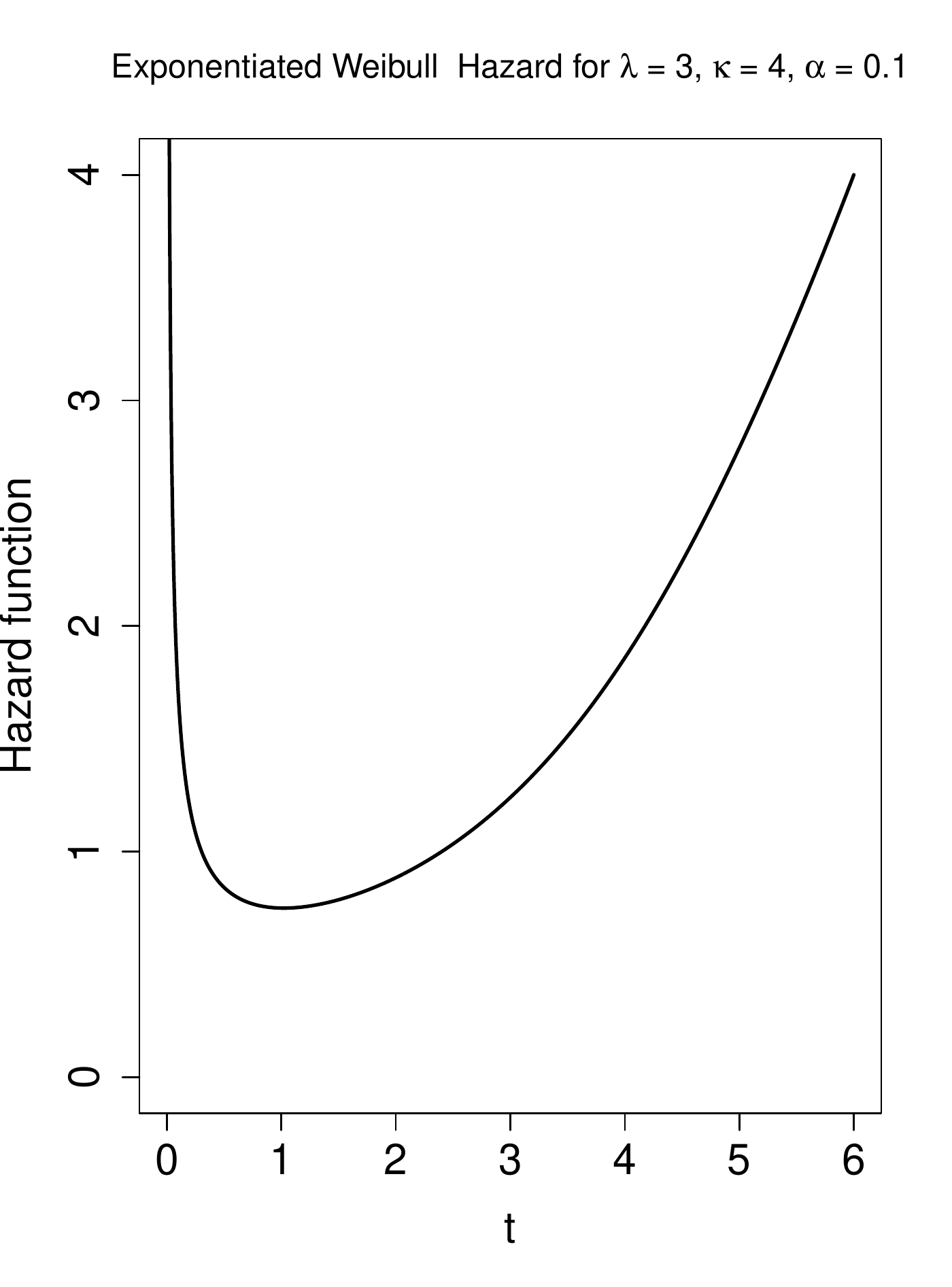}
\caption{Bathtub shape hazard for the exponentiated Weibull distribution for the event time t }
\end{figure}

\begin{figure}[bt]
\label{Figure:2}
\centering
\includegraphics[width = 11cm, height = 10cm]{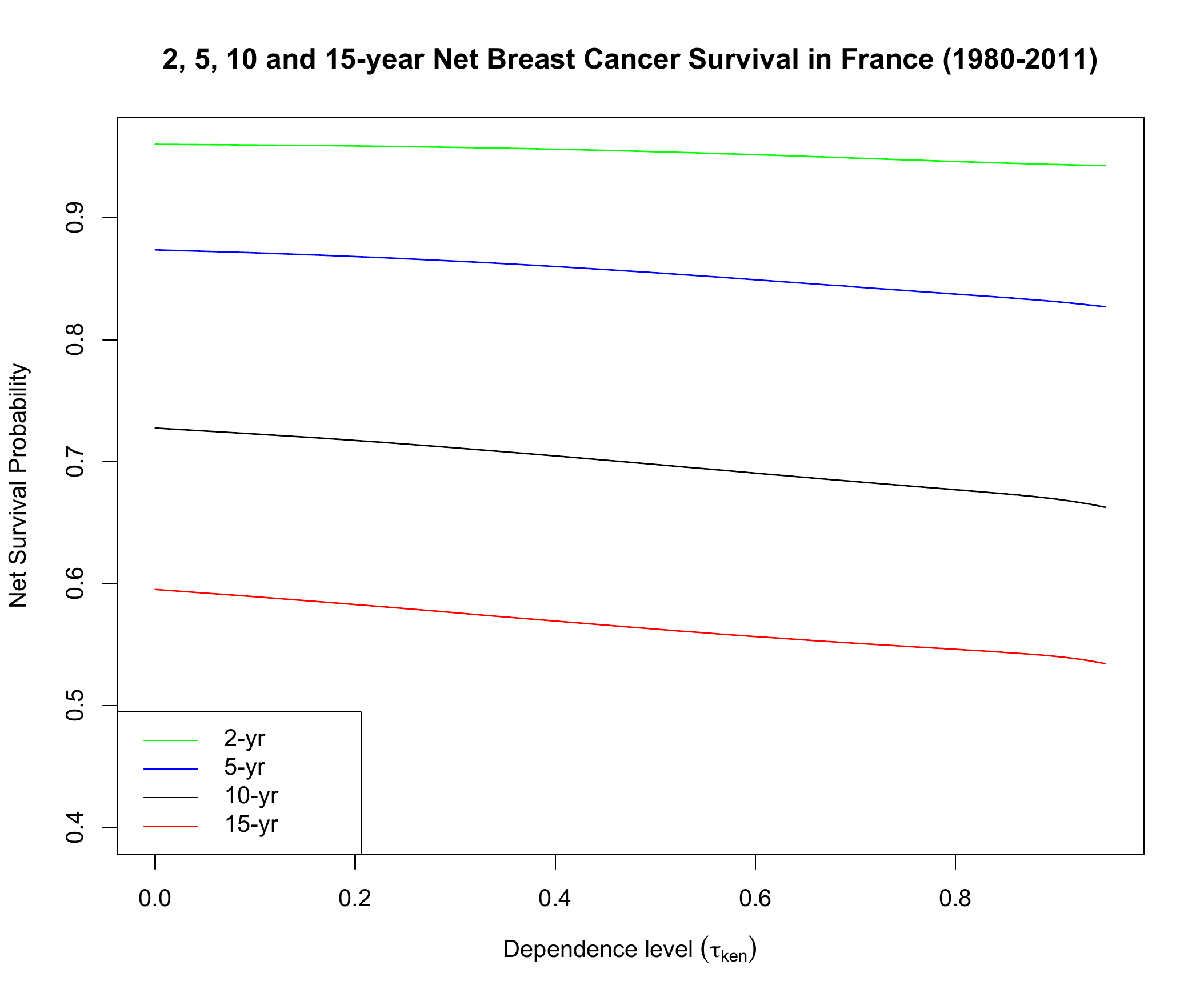}
\caption{Sensitivity Analysis of Net Breast Cancer Survival Across a Range of Dependent Competing Mortality }
\end{figure}

\begin{figure}[bt]
\label{Figure:3}
\centering
\includegraphics[width = 13cm, height = 11cm]{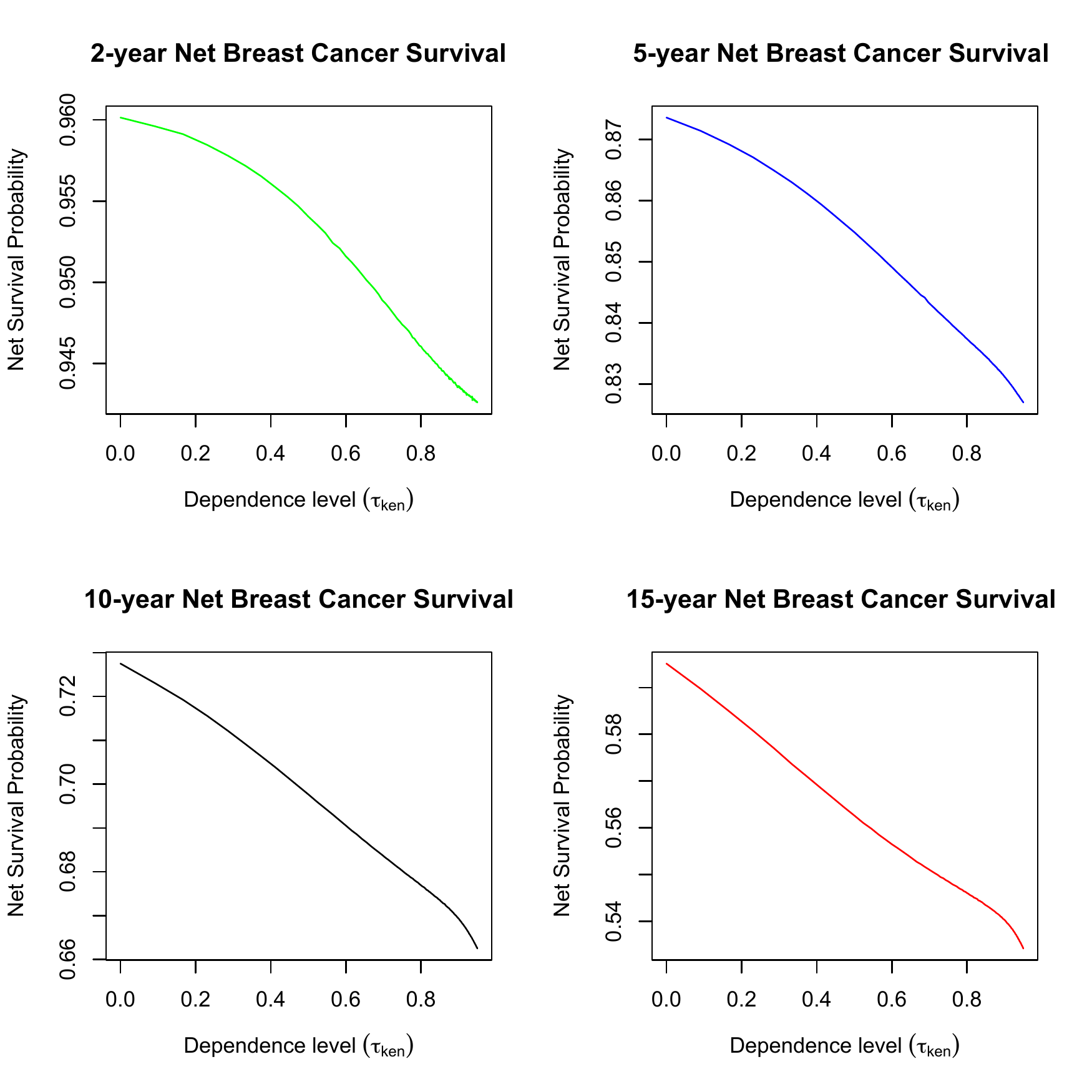}
\caption{Changes in Net Breast Cancer Survival for Increasing  Dependent Competing Risks}
\end{figure}
\end{document}